# The Efficacy of DO-178B

Dewi Daniels
Verocel Limited
Trowbridge, United Kingdom
ddaniels@verocel.com

*Abstract*— **DO-178B was based on the consensus of the avionic software community as it existed in 1992. Twenty two years after publication, we have no publically available experimental data as to its efficacy. It appears to work extremely well, since there have been no hull loss accidents in passenger service ascribed to software failure. This is a comforting and surprising result. However, if we don't know why DO-178B works so well, there is a danger that we could stop doing something that really matters, which could lead to an accident.**

## I. INTRODUCTION

I was greatly inspired by the book "The Machine That Changed the World" [1], which described research conducted in the late 1980s by the Massachusetts Institute of Technology (MIT) into best practice in the Japanese, US and European car industries as they existed at that time. I was impressed not only that the authors were able to identify fundamental differences in working practice between the most productive and the least productive car manufacturers, but that they were given access to detailed metrics that allowed them to quantify the result of following those working practices. The book concluded that lean production (a term coined by one of the MIT researchers) "uses less of everything compared with mass production – half the human effort in the factory, half the manufacturing space, half the investment in tools, half the engineering hours to develop a new product in half the time".

## II. SOFTWARE ENGINEERING

I am saddened that no published data exists for the efficacy of software engineering practices. The programming languages, tools and methodologies that are most widely adopted seem to be chosen on the basis of fads and fashions rather than any scientific evidence for their effectiveness or otherwise. Andy German [2] published the results of conducting static analysis on avionic software for the Lockheed C-130J. Andy concluded that "the poorest language for safety-critical applications is C with consistently high anomaly rates. The best language found is SPARK (Ada), which consistently achieves one anomaly per 250 lines of code". Nevertheless, weakly-typed languages continue to be used widely for software development.

One of the best books I have read on comparative software engineering practices is "Rise & Resurrection of the American Programmer" [3] by Ed Yourdon. Yourdon interviewed a number of organizations who he believed to exemplify best practice, including an Indian laboratory that was the first to achieve Capability Maturity Model (CMM) Level 5 and Microsoft, who were (and are) the World's best-selling software publisher. He found that Microsoft are rightly focused on maximizing product sales, not on minimizing their software development costs (which form only a small percentage of their total costs) or minimizing their software defect rate (since eliminating more defects could delay the introduction of a new product, potentially allowing a competitor to steal market share). It therefore follows that we should not look to the most successful software companies such as Microsoft for best practice when developing software for safety-critical applications. As it happens, I believe Microsoft to be the most improved software company of the last decade. For example, most Blue Screens of Death (BSODs) were not caused by Microsoft software, but by faulty third-party device drivers. The incidence of BSODs has been reduced dramatically by the deployment of tools based on formal methods, e.g. Static Device Verifier [4].

In the so-called soft sciences such as psychology and economics, I have noticed a sharp divide between experimental approaches that insist on the use of experiments to prove or disprove hypotheses, and theoretical approaches that develop complex theories from an initial set of axioms. For example, Robert Merton and Myron Scholes won the 1997 Nobel Prize in Economic Sciences for their work on the Black-Scholes equation. The failure by the financial community to understand the limitations of this model was one of the causes of the sub-prime lending crisis [5].

I worry that so little emphasis is placed on experiments in software engineering. I am also concerned there may be flaws in the little theory that we do have. For example, most software reliability models assume a normal (Gaussian) distribution of software defects. A software reliability model using a normal distribution assumes that software failures are independent of each other. This is not the case – software defects tend to be clustered. Mandelbrot published a book in 2004 [6] which predicted a stock market crash (which occurred in 2008) because the financial models assumed a normal distribution of share price movements. Mandelbrot claimed that share prices follow a power (fractal) distribution, not a normal distribution. The danger he cited was that a normal distribution would give the same prediction as a power distribution most of the time, but would considerably underestimate the probability of major fluctuations in the stock market.

## III. DO-178B

I am concerned that, 22 years after it was published, we have no statistical evidence as to the efficacy of DO-178B [7].

DO-178B appears to work very well. Despite the fact there are over 22,000 certified jet airplanes in service worldwide [8], no hull loss accidents in passenger service have been ascribed to software failure. I find this a comforting and surprising result.

The goal of DO-178B is to ensure that the system requirements allocated to software have been implemented correctly in the software. DO-178B defines five software levels, from Level A software, which could cause or contribute to a failure of system function resulting in a catastrophic failure condition for the aircraft, to Level E software, whose failure would have no effect on aircraft operational capability or pilot workload. There is no publically available evidence of the defect rate achieved by DO-178B Level A, let alone any difference between the five software levels or the efficacy of any of the individual DO-178B objectives. Andy German [2] reported "when Level A was compared to Level B, no significant difference in anomaly rates identified by static analysis was found".

Requirements-based testing is a strength of DO-178B. Peter Ladkin's compendium of computer-related incidents with commercial aircraft [9] shows that in the majority of incidents (e.g. the Lufthansa A320 runway over-run at Warsaw), the software faithfully implemented requirements that specified unsafe behavior under some unforeseen circumstance. Even DO-178B Level A software is far from defect-free, but there seem to have been very few incidents caused by software failing to satisfy its requirements.

John Rushby [10] wrote, "the standards-based approach to certification employed with DO-178B does seem to be effective, and this is probably because its prescriptions are based on experience and are sound, and because they are executed diligently and monitored conscientiously – but also perhaps because of factors outside the standards relating to safety culture, experience and conservatism". John worries that "increased outsourcing and other changes in the aircraft industry reduce some factors that may, implicitly, have contributed to the safety of aircraft software (e.g. organizational experience and safety culture)".

I worry that we have previously relied less on automation than we have supposed. For example, when an Air Data Inertial Reference Unit (ADIRU) failure caused an in-flight-upset on a Boeing 777 [11], the pilot was able to disengage the autopilot and land normally. There seems to be growing evidence (e.g. Colgan Air Flight 3407, Air France 447) that crews are becoming less able to cope with failures of automation, meaning that we need to improve pilot training, improve the reliability of the automation or accept that some accidents will happen when the automation fails.

## IV. PROPOSED EXPERIMENT

It would be helpful to determine the efficiency of the five DO-178B software levels by measuring the number of defects found in relation to the software verification effort expended.

The ideal would be a study that used historical data to assess the in-service history of software developed to DO-178B. It is considered unlikely that such data will be made available by industry or that it even exists.

An achievable experiment would be to take some pre-existing software and to verify that software to DO-178B Levels A, B, C and D. A suitable candidate would be open source software such as CUnit or OpenSSL. We suggest that one team could verify the software to Level D, while a second team could verify the software to Level C, then carry out the additional work necessary to meet the objectives of Level B and Level A in turn.

## V. CONCLUSION

Those seeking to reduce costs argue that some of the DO-178B objectives or activities are unnecessary and could be eliminated. The danger is that, if we don't know why DO-178B works, we could stop doing something that really matters, which could lead to an accident.